\documentclass[12pt,a4paper]{scrartcl}
\pdfoutput=1
\usepackage{amsmath,amssymb}
\usepackage{subfigure}

\usepackage{hyperref}
\usepackage{graphicx}

\newcommand{\hhref}[1]{\href{http://arxiv.org/abs/#1}{arXiv:#1}}

\addtokomafont{disposition}{\rmfamily\boldmath}
\let\tmptitle\title\renewcommand{\title}[1]{\tmptitle{\LARGE #1}}
\let\tmpauthor\author\renewcommand{\author}[1]{\tmpauthor{\large #1}}
\let\tmpdate\date\renewcommand{\date}[1]{\tmpdate{\normalsize #1}}
\newcommand{\abstrct}[1]{\begin{abstract}\vspace{-2em}\small\noindent#1\end{abstract}}

\title{\Large
Neutrino Masses and Leptogenesis from Extra Fermions
}
\date{\today}
\author{
Dmitry~V.~Zhuridov
\footnote{Email: dmitry.zhuridov@wayne.edu}
\\ \normalsize\itshape
Scuola Normale Superiore, Piazza dei Cavalieri 7, 56126 Pisa, Italy
\\ \normalsize\itshape
Department of Physics and Astronomy, Wayne State University, Detroit, MI 48202, USA
}

\begin{document}

\maketitle

\abstrct{
Generation of the neutrino masses and leptogenesis (LG) in the Standard Model extended by the heavy 
Majorana fermions is considered. Classification of LG scenarios according to the new fermion mass spectra is given, where singlet-triplet LG is considered for the first time. The upper bound on the $CP$ asymmetry relevant for LG with hierarchical heavy neutrinos (Davidson-Ibarra bound) is revised, and shown that in the case of one massless neutrino it essentially depends on the type of the light neutrino mass hierarchy.
The resonant scenarios, which help to avoid the problem of extremely high reheating temperature in the early universe, are discussed. 
In particular, we present new simplified, generalized and detailed formulation of freed LG, which violates Davidson-Ibarra bound in a special class of models.
}

\section{Introduction}

The observable small neutrino masses and large baryon asymmetry of the Universe~\cite{PDG2010} can be economically explained by the 
see-saw mechanism~\cite{seesawA}-\cite{seesawF} and the baryogenesis~\cite{Sakharov,KRS} via leptogenesis (LG) scenario~\cite{Fukugita_Yanagida}, 
respectively, which are widely discussed in the literature. So far many variations of this mechanisms have been found with 
different new fermion and scalar representations involved, where the neutrino masses can be generated on the tree and loop levels~\cite{Babu:2001ex}-\cite{Mazumdar:2012qk}, while LG occurs through the out-of-equilibrium decays of the heavy fields involved in the see-saw.

According to Sakharov criteria~\cite{Sakharov}, one of the central roles in these theories is played by the $CP$ asymmetry $\epsilon$ relevant for LG. 
In the case of  {\it hierarchical} heavy neutrinos the model-independent upper limit on $|\epsilon|$ is proportional to the light neutrino mass difference, and called Davidson-Ibarra bound~\cite{DI0A}-\cite{Gorbunov:2011zz}, which implies strong lower bound on the masses of the heavy neutrinos involved in LG. Thermal production of such heavy neutrinos in the early universe requires high reheating temperature, which may lead to overproduction of light states possibly present in the spectrum of a theory. This is called a gravitino problem~\cite{Khlopov1,Balestra} since in the supersymmetric models such light states are typically gravitinos. 

However $\epsilon$ may be resonantly enhanced, which helps to avoid the gravitino problem. Popular mechanism of such enhancement due to possible {\it quasi-degeneracy} of heavy neutrino masses is known as resonant LG~\cite{resonantLG1}-\cite{resonantLG6}, and may take place even at TeV scale. Moreover it was recently discovered in Refs.~\cite{KZ,freedLG} that even for {\it hierarchical} heavy neutrinos Davidson-Ibarra bound can be violated in the class of models with partial cancellation between various (e.g., tree and loop) contributions to the observable neutrino masses. The mechanism with this new type of resonant enhancement of $\epsilon$ (expressed in terms of light neutrino masses) is called {\it freed LG} since it significantly frees the allowed parameter space for successful LG, and may improve testability of the underlying models.  In this way, freed LG extends the benefits of resonant LG to more general case of hierarchical heavy neutrinos.


In this paper we present new simplified, generalized and detailed formulation of freed LG. 
To start with, in the next section we consider standard model (SM) extensions by heavy Majorana fermions, which possess the see-saw mechanism. 
In these extensions the neutrino masses can be generated by the tree level exchange of the fermion singlets, triplets and both singlets plus triplets, which is called type I~\cite{seesawA}-\cite{seesawF}, III~\cite{seesawIIIA}-\cite{seesawIIIC} and hybrid I+III~\cite{Barr:2005je}-\cite{Perez:2007iw} see-saw mechanism, respectively. 

The classification and uniform description of various  types of LG in the considered models we present in section~\ref{section:fermionicLG}. 
Among them there are four basic types, since the decaying lightest non-SM fermion and the heavier fermion, which propagates in the loop diagrams, may be either $SU(2)$ singlet or triplet. 
All these possible mechanisms we call {\it fermionic LG}. 
Pure singlet and pure triplet LG were described uniformly in Ref.~\cite{Strumia}. Triplet-singlet LG~\cite{LG_BP1,LG_BP2} arises naturally in 
theories of grand unification, such as Adjoint $SU(5)$ model~\cite{AdjointSU5}, in which the triplet should be lighter than singlet to 
guarantee the unification. The last possibility of singlet-triplet LG we consider for the first time. This mechanism can be realized, e.g., in SUSY $SU(5)$ model, where the exact masses of triplet and singlet fermions are not determined from phenomenological constraints~\cite{SUSY_SU(5)}.

In section~\ref{section:Hierarchical case} we have revised the derivation of Davidson-Ibarra bound and pointed out that in the models with one massless neutrino it essentially depends on the type of the hierarchy of light neutrino masses.
The possibility to weaken the bound on $\epsilon$ is briefly discussed for nearly degenerate heavy neutrinos in section~\ref{section:Resonant LG}, and considered in more details for hierarchical heavy neutrinos in the following, in particular, general explanation of freed LG is presented in section~\ref{section:FreedLG} with more details given in the Appendix~\ref{app:CPasymmetry}, and its possible realization in Adjoint $SU(5)$ model is demonstrated in section~\ref{section:FreedLGinAd}.
And we conclude in section~\ref{section:summary}.

\section{Neutrino masses}\label{section:nu_masses_generation}

Consider the see-saw generation of the neutrino masses in the SM extended by extra heavy fermions, see Ref.~\cite{Davidson:2008bu} for a good review. 
For uniform description we add to the SM Lagrangian the {\it singlet} Majorana fermion terms
\begin{eqnarray}\label{eq:Lsinglet}
 {\cal L}_\text{singlet} = \bar N_i \text{i}\gamma^\mu \partial_\mu N_i - \left( Y_{i\alpha}^sN_i\bar L_\alpha\phi 
+ \frac{1}{2}\bar N_iM_{ij}^sN_j^c + \text{H.c.} \right)
\end{eqnarray}
or/and $SU(2)_L$ {\it triplet} Majorana fermion terms
\begin{eqnarray}\label{eq:Ltriplet}
 {\cal L}_\text{triplet} = \text{Tr}\bar T_i \text{i}\gamma^\mu D_\mu T_i - \left( Y_{i\alpha}^t\bar L_\alpha T_i\phi 
+ \text{Tr}\bar T_iM_{ij}^tT_j^c + \text{H.c.} \right),
\end{eqnarray}
where $\alpha=e,\mu,\tau$ is a flavor index, $i$ and $j$ numerate the new fermions, the summation over repeated indexes and proper contraction 
under $SU(2)$ is assumed, $c$ stands for charge conjugation, the matrix representation for $T_i$ is given by
\begin{eqnarray}
T_i = \frac{1}{2}
 \begin{pmatrix}
  T_i^0 & \sqrt{2}T_i^+ \\
  \sqrt{2}T_i^- & -T_i^0
 \end{pmatrix},
\end{eqnarray}
the long derivative acts as
\begin{eqnarray}
 D_\mu T_i=\partial_\mu T_i+\text{i}g_2[W_\mu,T_i],
\end{eqnarray}
and we use the notation $L=(e_L,\nu_L)^T$ and $\phi=(\phi^0,\phi^-)^T$  for the SM lepton and Higgs doublets, respectively.

Integrating out the heavy fermions in Eqs.~\eqref{eq:Lsinglet} and \eqref{eq:Ltriplet} we get the standard Weinberg dimension 5 operator~\cite{Weinberg:1979sa}
\begin{eqnarray}\label{eq:eff_operator_Y}
 	{\cal L}_Y^\text{eff}  = \frac{1}{2} \bar L\phi \frac{M_\nu^\text{tree}}{v_0^2} \phi^T L^c + \text{H.c.}
\end{eqnarray}
with $3\times3$ neutrino mass matrix 
\begin{eqnarray}
 M_\nu^\text{tree} = v_0^2 \left[ Y^{sT}(M^s)^{-1}Y^s + Y^{tT}(M^t)^{-1}Y^t \right],  \label{eq:nu_masses:0}
\end{eqnarray}
where $v_0=174$\,GeV is the Higgs vacuum expectation value. 
We assume that the total number of the new fermions is $n=2$ or 3, and choose the bases for $N_i$, $T_i$ and charged leptons where the mass matrices $M^s$, 
$M^t$ and the charged lepton Yukawa matrix are diagonal and real. Hence in the hybrid case with both singlet and triplet fermions 
one of $Y^s$ and $Y^t$ is a complex $(n-1)\times3$ matrix (or row), while the other one is a row of 3 complex elements. 

In the following we use sequentially numbered singlets and triplets 
in ascending order of their masses $0<M_1<M_2<\dots$. 
Hence the neutrino mass matrix in Eq.~\eqref{eq:nu_masses:0} can be rewritten as
\begin{eqnarray}
 M_\nu^\text{tree} = v_0^2 Y^TD_M^{-1}Y,  \label{eq:nu_masses:1}
\end{eqnarray}
where $D_M=\text{diag}(M_1,M_2,\dots)$ and $Y$ is a general complex $n\times3$ matrix. In the pure case with only singlet 
(triplet) fermions the matrix $Y$ coincides with $Y^s$ ($Y^t$), while in the hybrid case it is constructed from $Y^s$ and $Y^t$ by 
adding to the $(n-1)\times3$ matrix (or row) the row in the proper place according to the index number of the corresponding mass eigenstate. 
The 3 phases can be removed from $Y$ by phase redefinition on the charged leptons. For $n=3$ this leaves 9 moduli and 6 phases as physical parameters, 
similarly to the type~I see-saw case~\cite{Davidson:2008bu}, while for $n=2$ only 6 moduli and 3 phases are left.
  
The lightness of the observable neutrinos can be explained by the small ratio $v_0/D_M$ in Eq.~\eqref{eq:nu_masses:1}, which is the usual see-saw factor. We recall that for $n=2$ the lightest neutrino is massless.

\section{Leptogenesis}\label{section:LG}
\subsection{Fermionic LG}\label{section:fermionicLG}

Consider the standard fermionic LG, in which the $CP$ violation is generated in the out-of-equilibrium decays of heavy Majorana fermions through 
the interference of tree and one-loop diagrams. Clearly it is important 
whether the decaying fermion is singlet or triplet, and same for the fermion, which propagates in the loop diagrams. 
According to this consider the four types of spectrum, shown in the Table~\ref{table:1}, where and below $i=1,2,\dots$ and $j=2,\dots$ 
\begin{table}[htdp]
\caption{Characteristics of the four basic types of fermionic LG. }
\label{table:1}
\begin{center}
\begin{tabular}{|c|c|c|c|c|c|c|}
	\hline
    \# & Type of LG & New fermions & $F_j$ & $F$ & $M_1^{\min}$ (resonant LG) \\ 
    \hline
    1 & Pure singlet & $\{N_i\}$ & $\frac{2S_j+V_j}{3}$ & 1 & well below 1\,TeV \\ 
    2 & Pure triplet & $\{T_i\}$ & $\frac{2S_j-V_j}{3}$ & 1/3 & $\sim1.6$\,TeV  \\
    3 & Singlet-Triplet & $N_1$, $\{T_j\}$ & $V_j$ & 1 & --- \\ 
    4 & Triplet-Singlet & $T_1$, $\{N_j\}$ & $\frac{V_j}{3}$ & 1/3 & --- \\  
    \hline
\end{tabular}
\end{center}
\label{default}
\end{table}

For all the listed cases the $CP$ asymmetry can be expressed uniformly (see Ref.~\cite{Strumia} for the cases 1 and 2, and Ref.~\cite{LG_BP1} for the case 4)
\begin{eqnarray}\label{eq:eps1}
 \epsilon_1 = -\sum_j\frac{3}{2} F_j\frac{M_1}{M_j} \frac{\Gamma_j}{M_j} I_j,
\end{eqnarray}
where we summed over the lepton flavors\footnote{For simplicity we do not consider the flavor effects~\cite{Davidson:2008bu,Abada:2006fw,Nardi:2006fx}, which may change the result by factor of few, since this is a minor effect for our main discussion in sections \ref{section:FreedLG} and \ref{section:FreedLGinAd}.},
\begin{eqnarray}
 I_j = \frac{\text{Im}[(YY^\dag)_{1j}^2]}{(YY^\dag)_{11} (YY^\dag)_{jj}}
\end{eqnarray}
and
\begin{eqnarray}
 \frac{\Gamma_j}{M_j} = \frac{(YY^\dag)_{jj}}{8\pi}
\end{eqnarray}
with the total width $\Gamma_j\equiv \Gamma(N_j\to \ell\phi,\bar\ell\bar\phi)$ of $N_j$ decay into a two particle final state.

The expressions of the loop factor $F_j$, which are shown in Table~\ref{table:1} for the discussed types of LG, contain the self energy
\begin{eqnarray}\label{eq:S-factor}
 S_j = \frac{M_j^2\Delta M_{1j}^2}{(\Delta M_{1j}^2)^2+M_1^2\Gamma_j^2}
\end{eqnarray}
and vertex
\begin{eqnarray}\label{eq:V-factor}
 V_j = 2\frac{M_j^2}{M_1^2} \left[ \left(1+\frac{M_j^2}{M_1^2}\right) \log\left(1+\frac{M_1^2}{M_j^2}\right) - 1 \right]
\end{eqnarray}
loop terms, where
\begin{eqnarray}
 \Delta M_{ij}^2 = M_j^2 - M_i^2.
\end{eqnarray}

In the case of other possible types of the extra fermion spectrum one should take the proper combination of the 
$CP$ asymmetries, namely, a superposition of pure singlet and singlet-triplet (pure triplet and triplet-singlet) expressions should be taken 
in the singlet-triplet-singlet (triplet-singlet-triplet) case. 

Interestingly, the self energy contribution vanishes in the hybrid cases 3 and 4. 
It was found in Ref.~\cite{LG_BP1} for the case 4, and we get similar cancellation in the case 3.  
The additional factor 3 in the expression for $F_j$ in the singlet-triplet case comparing to the triplet-singlet one is related to the 3 
components of triplet running in the vertex loop.

\subsection{Hierarchical case}\label{section:Hierarchical case}

The loop factors in Eqs.~\eqref{eq:S-factor} and \eqref{eq:V-factor} are defined in such way that
in the hierarchical limit $M_j/M_1\to\infty$ they go to unity: $S_j=1$ and $V_j=1$. The $CP$ asymmetry in this limit can be written as
\begin{eqnarray}\label{eq:epsilon-1}
 \epsilon_1 = -\frac{3F}{16\pi} \frac{M_1\Sigma_\nu^\text{tree}}{(YY^\dag)_{11}},
\end{eqnarray}
where the values of $F=\lim_{M_j/M_1\to\infty} F_j$ are shown in Table~\ref{table:1}, and 
\begin{eqnarray}\label{eq:Sigma}
 \Sigma_\nu^\text{tree} = \sum_j \text{Im}[(YY^\dag)_{1j}^2M_j^{-1}] =  \frac{1}{v_0^2} \text{Im}\left\{ \left[ Y\left(M_\nu^\text{tree}\right)^\dag 
Y^T \right]_{11} \right\}.
\end{eqnarray} 
Using the Casas--Ibarra parametrization~\cite{Casas-Ibarra}
\begin{eqnarray}\label{eq:Casas-Ibarra}
 Y = v_0^{-1}D_{M}^{1/2}R D_\nu^{1/2}U^\dag,
\end{eqnarray}
where $U$ is the PMNS leptonic mixing matrix~\cite{PMNS1,PMNS2}, which diagonalizes the neutrino mass matrix 
in the flavor basis (in which the charged-lepton Yukawa matrix and gauge interactions are flavor-diagonal) as
\begin{eqnarray}\label{eq:PMNS}
  U^T M_\nu^\text{tree} U = {\rm diag}(m_1,m_2,m_3) \equiv D_\nu,
\end{eqnarray}  
and $R$ is a complex orthogonal matrix (or partly orthogonal in the case 
$m_1=0$~\cite{Ibarra-Ross}),
the $CP$ asymmetry can be rewritten as
\begin{eqnarray}\label{eq:CPasymmetry0}
 \epsilon_1^0 = -\frac{3F}{16\pi} \frac{M_1}{v_0^2} \frac{\sum_k m_k^2\, \text{Im}[ R_{1k}^2]}{\sum_k m_k|R_{1k}|^2}.
\end{eqnarray} 
In the notation $R_{1k}=x_k+\text{i}y_k$ it takes the form
\begin{eqnarray}\label{eq:CPasymmetry1}
 \epsilon_1^0 = -\frac{3F}{16\pi} \frac{M_1}{v_0^2} \frac{2\sum_k m_k^2 x_k y_k}{\sum_k m_k(x_k^2+y_k^2)}
\end{eqnarray}
with the orthogonality condition $\sum_k (x_k+\text{i}y_k)^2 = 1$.

For $0<m_1<m_2<m_3$ the upper bound is
\begin{eqnarray}\label{eq:CPasymmetryMAX}
	 |\epsilon_1|^\text{max} = \frac{3F}{16\pi} \frac{M_1}{v_0^2} (m_3-m_1)
\end{eqnarray}
with the choice of $\vec{x}=x(\pm1,0,1)$ and $\vec{y}=y(\mp1,0,1)$, where $x^2=y^2+1/2 \gg 1$. However this choice is forbidden  
in the case with one massless neutrino, in which $R$ is $2\times3$ matrix with zero first column~\cite{Ibarra-Ross}
\begin{eqnarray}
 R = 
\left(
 \begin{array}{ccc}
  0 & \cos z & \pm\sin z \\
  0 & -\sin z & \pm\cos z \\
 \end{array}
\right),
\end{eqnarray}
where $z=\alpha+\text{i}\beta$ is the complex angle. By using this $R$, Eq.~\eqref{eq:CPasymmetry0} can be rewritten as
\begin{eqnarray}\label{eq:CPasymmetry0alpha}
 	\epsilon_1^0 = -\frac{3F}{16\pi} \frac{M_1}{v_0^2} \frac{(m_3^2-m_2^2)\sin2\alpha\sinh2\beta}{(m_2+m_3)\cosh2\beta-(m_3-m_2)\cos2\alpha}.
\end{eqnarray} 
The maximization of the trigonometric factor in Eq.~\eqref{eq:CPasymmetry0alpha} over $\alpha$ gives
\begin{eqnarray}
 	-\frac{3F}{16\pi} \frac{M_1}{v_0^2}  (m_3-m_2) \tanh2\beta.
\end{eqnarray} 
Hence
for $0=m_1<m_2<m_3$ we have
\begin{eqnarray}\label{eq:CPasymmetryMAX2}
 |\epsilon_1|^\text{max} = \frac{3F}{16\pi} \frac{M_1}{v_0^2} (m_3-m_2).
\end{eqnarray}
The neutrino mass difference in Eq.~\eqref{eq:CPasymmetryMAX} is $m_3-m_1\approx\sqrt{|\Delta m_\text{atm}^2|}$ for both normal and inverted hierarchies of the neutrino masses.\footnote{For quasi-degenerate neutrino masses $m_1\approx m_2\approx m_3\approx m_0$ we have $m_3-m_1\approx |\Delta m_\text{atm}^2|/(2m_0)$.} However the difference $m_3-m_2$ in Eq.~\eqref{eq:CPasymmetryMAX2} is about $\sqrt{|\Delta m_\text{atm}^2|}$ only for the case of the normal hierarchy, while for the inverted one the result is $m_3-m_2\approx\sqrt{\Delta m_\text{sol}^2}/2$, which suppresses $|\epsilon_1|^\text{max}$ by 1 order of magnitude.

From this Davidson--Ibarra bound on $\epsilon_1$~\cite{DI0A}-\cite{Davidson:2002qv}, using $m_3^2\approx |\Delta m_\text{atm}^2|=2.40\times10^{-3}~\text{eV}^2$~\cite{PDG2010} 
and the baryon asymmetry of the Universe $\eta_B\approx6\times10^{-10}$, one can get the bound~\cite{Davidson:2002qv,Chen:2007fv}
\begin{eqnarray}\label{eq:boundM1}
M_1\gtrsim10^9\, \text{GeV},
\end{eqnarray}
which would be even stronger for either quasi-degenerate light neutrinos or inverted hierarchical neutrinos with one massless state.
However this bound can be significantly relaxed in freed LG~\cite{KZ,freedLG}, which is discussed in section~\ref{section:FreedLG}.

\subsection{Resonant LG}\label{section:Resonant LG}

The bound in 
Eqs.~\eqref{eq:CPasymmetryMAX} and \eqref{eq:CPasymmetryMAX2} does not make sense for nearly 
degenerate heavy Majorana fermions in the resonant LG~\cite{resonantLG1}-\cite{resonantLG6}, 
which is possible at TeV scale. Indeed in the limit $M_j-M_1\ll M_1$ 
the self energy diagrams dominate. Using Eq.~\eqref{eq:S-factor}, for $M_j-M_1\approx\Gamma_j/2$ we have $S_jM_1\Gamma_j/M_j^2\approx1$. 
Hence $|\epsilon_1|\sim1$ in Eq.~\eqref{eq:eps1} for $I_j\sim1$.
We note that this resonant scenario is not applicable to the singlet-triplet and triplet-singlet LG, which do not have the self-energy 
contribution in the $CP$ asymmetry,  
as shown in Table~\ref{table:1}. 

The scattering term in the Boltzmann equation for a triplet fermion number density has gauge boson mediated contribution 
in addition to the Higgs mediated one, which is presented in both singlet and triplet fermion cases. In result, TeV triplets are thermalized by 
gauge boson mediated annihilations up to $M_1/T\gg1$, where $T$ is the temperature. The efficiency factor in the final baryon asymmetry 
depends in this case on the triplet mass and is strongly suppressed for $M_1\sim\mathcal{O}(\text{TeV})$~\cite{1007.1907}. 
This implies $M_1\gtrsim1.6$\,TeV~\cite{0806.1630,Burnier:2005hp} in contrast to the lower bound on the fermionic singlet mass, which can be well below the TeV 
scale.
Table~\ref{table:1} shows the lower bounds on $M_1$ for resonant LG.

\subsection{Freed LG}\label{section:FreedLG}

Using Eqs.~\eqref{eq:eff_operator_Y} and \eqref{eq:nu_masses:1}, the Weinberg operator can be rewritten as
\begin{eqnarray}\label{eq:eff_operator_Y1}
 	{\cal L}_Y^\text{eff}   = \frac{1}{2} \bar L\phi \frac{M_\nu^\text{tree}}{v_0^2} \phi^T L^c + \text{H.c.}
	= \frac{1}{2} \bar L\phi(Y^T D_M^{-1}Y)\phi^T L^c + \text{H.c.}
\end{eqnarray}
In this section we discuss the new kind of LG, which takes place in the theories with the additional to Eq.~\eqref{eq:eff_operator_Y1} effective dim.5 operator\footnote{This operator can be generated in loops with the additional to $N_i$ or/and $T_i$ new particles (including scalars) in the theory. However they do not participate directly in generation of the $CP$ asymmetry in freed LG, which therefore can be attributed to the class of fermionic LG.}
\begin{eqnarray}\label{eq:eff_operator}
	 {\cal L}_h^\text{eff} = -\frac{1}{2\Lambda} \bar L_\alpha\phi(h_\alpha h_\beta^T)\phi^T L_\beta^c + \text{H.c.},
\end{eqnarray}
where $\Lambda>0$ is the high-energy scale.
The operator in Eq.~\eqref{eq:eff_operator} produces new contribution $-hh^Tv_0^2/\Lambda$ to the neutrino mass matrix besides the one in Eq.~\eqref{eq:nu_masses:1}. 

In many theories of grand unification, e.g., in Adjoint $SU(5)$ model~\cite{AdjointSU5}, the new Yukawa couplings $h$ are proportional to $Y$~\cite{KZ},~\cite{freedLG}
\begin{eqnarray}\label{eq:linear_relation}
 h_\alpha^T = a_iY_{i\alpha}.
\end{eqnarray}
(Here we consider real $a_i$.) 
In this case, by using the orthogonal transformation $QY=Y^\prime$, the neutrino mass matrix 
can be rewritten in the form of Eq.~\eqref{eq:nu_masses:1} as
\begin{eqnarray}\label{eq:nu_masses:2}
 	M_\nu^\text{tree} - v_0^2\frac{Y^Ta^TaY}{\Lambda}  \equiv   M_\nu = v_0^2\, Y^{\prime T} D_\mathcal{M}^{-1}Y^{\prime}
\end{eqnarray} 
with the diagonal modified mass matrix of heavy fermions
$D_\mathcal{M} = \text{diag}(\tilde M_{1},\tilde M_{2},\dots)$~\cite{freedLG}.

Note that the relative minus sign between Eqs.~\eqref{eq:eff_operator_Y1} and \eqref{eq:eff_operator} 
may result in partial cancellation between the ordinary see-saw and the new contribution to the neutrino mass matrix in Eq.~\eqref{eq:nu_masses:2}, which relaxes  the $CP$ asymmetry in LG, expressed through the light neutrino masses, as discussed below. This minus sign can be generated, e.g., due to negative new scalar coupling in the potential, such as $\lambda_5$ in the next section~\cite{KZ,freedLG}.

For the case of two new fermions in the hierarchical limit $M_1/M_2\to0$ 
the sum $\Sigma_\nu^\text{tree}$ in Eqs.~\eqref{eq:epsilon-1} and \eqref{eq:Sigma} can be rewritten as $\Sigma_\nu^\text{tree}=\mu\Sigma_\nu$ with the magnification factor 
$\mu=\Lambda/(\Lambda-a_2^2M_2)$~\cite{freedLG} (see appendix~\ref{app:CPasymmetry} for the details).
Hence the $CP$ asymmetry is proportional to $\mu$ and can be resonantly enhanced. In the limit of $M_1\ll \text{min}(\Lambda, M_2)$, using the Casas--Ibarra parametrization of $Y^\prime$ with the matrix $R^\prime$ in Eq.~\eqref{eq:Casas-Ibarra^prime}, we have 
\begin{eqnarray}\label{eq:epsilon_freed1}
 \epsilon_1^{\text{freed}}  = -\frac{3F}{16\pi} \frac{\mu M_1}{v_0^2} \frac{\sum_k m_k^2 \text{Im} 
[R_{1k}^{\prime2}]}{\sum_k m_k|R_{1k}^\prime|^{2}} 
\end{eqnarray}
for $a_1^2a_2^2M_1M_2\ll \Lambda|\Lambda-a_2^2M_2|$ case, which is natural, but implies the restriction $|\mu|\ll \Lambda^2/(a_1^2a_2^2M_1M_2)$, in particular, $|\mu|\ll \Lambda/(a_1^2M_1)$ for large $|\mu|$; and
\begin{eqnarray}\label{eq:epsilon_freed2}
 \epsilon_1^\text{freed} = -\frac{3F}{16\pi} \frac{\mu M_1}{v_0^2} \frac{\sum_k m_k^2 \text{Im}[ R_{2k}^{\prime2}]}{\sum_k m_k|R_{2k}^\prime|^{2}} 
 \, \text{sign}[\Lambda-a_2^2M_2]
\end{eqnarray}
for $a_1^2a_2^2M_1M_2\gg \Lambda|\Lambda-a_2^2M_2|$ case, which requires strong fine tuning. 

Both expressions~\eqref{eq:epsilon_freed1} and \eqref{eq:epsilon_freed2} generate the upper bound of the same form 
\begin{eqnarray}
	|\epsilon_1^\text{freed}|^\text{max} = |\mu||\epsilon_1|^\text{max},
\end{eqnarray}
where $|\epsilon_1|^\text{max}$ is given in Eqs.~\eqref{eq:CPasymmetryMAX} and \eqref{eq:CPasymmetryMAX2}.
Note that in the intermediate region of $a_1^2a_2^2M_1M_2\sim \Lambda|\Lambda-a_2^2M_2|$ the approximate relation $|\epsilon_1^\text{freed}|^\text{max} \approx |\mu||\epsilon_1|^\text{max}$ still takes place.

In the resonant case $\Lambda\approx a_2^2M_2$ we have $|\mu|\gg1$, which enhances the $CP$ asymmetry, 
and gives more freedom to the mass of heavy neutrino $N_1$ by relaxing the bound in Eq.~\eqref{eq:boundM1} as
\begin{eqnarray}
M_1\gtrsim 10^9|\mu|^{-1}\, \text{GeV}.
\end{eqnarray}

\subsection{Freed LG in Adjoint $SU(5)$}\label{section:FreedLGinAd}

One of the characteristic examples for possible realization of freed LG~\cite{KZ,freedLG} is Adjoint $SU(5)$ model~\cite{AdjointSU5}, which 
is a viable grand unified theory with the massive neutrinos and suitable proton lifetime. 
In this model, the ordinary $SU(5)$ theory~\cite{Georgi:1974sy} particle content is extended by the new fermions in adjoint $24_F$ representation and bosons in $45_H$. In result, in addition to the tree level III+I seesaw neutrino mass term there is a one-loop contribution with octet fermions $\rho_8\in24_F$ and scalars $S_8\in45_H$ propagating in the loop. The lightest new fermion, which plays role of $T_1$ in the above sections of this paper, is the fermionic triplet $\rho_3\in24_F$. The linear coefficients $a_1$ and $a_2$ in Eq.~\eqref{eq:linear_relation} are determined in Adjoint $SU(5)$ by the ratio of the vacuum expectation values of the SM Higgs doublet $\phi$ (mixture of the two scalar doublets in $5_H$ and $45_H$) and $45_H$.

Fig.~1 shows the dependence of the magnification factor $\mu$ on the absolute value of $a_2$ for the chosen values of $S_8$ mass, masses $M_{\rho_3}=5\times10^{11}$~GeV and $M_{\rho_8}=10^{15}$~GeV of triplet and octet adjoint fermions, respectively, and the coupling $\lambda_5=\{-0.5,-0.1\}$ of the interaction $(\phi^\dag S_8)^2$. 
In this choice of the parameter values we took into account the proton decay constraint on scalar octet mass $M_{S_8}<4.4\times10^{-5}$~GeV, 
the regime of successful unification constraint on the adjoint fermion mass ratio $M_{\rho_8}/M_{\rho_3}>100$, 
and the unflavored LG (without usage of any resonance conditions) bound on triplet fermion mass $M_{\rho_3}\gtrsim5\times10^{11}$~GeV.

Fig.~1 demonstrates that for essential range of the allowed parameter space the value of $\mu$ is significantly larger than unity, which may enhance the $CP$ asymmetry in LG and relax the allowed parameter space of the model, in particular, the lower bounds on $M_{\rho_3}$ and other adjoint fermions.

\begin{figure}
  \centering
  \includegraphics[width=.45\textwidth]{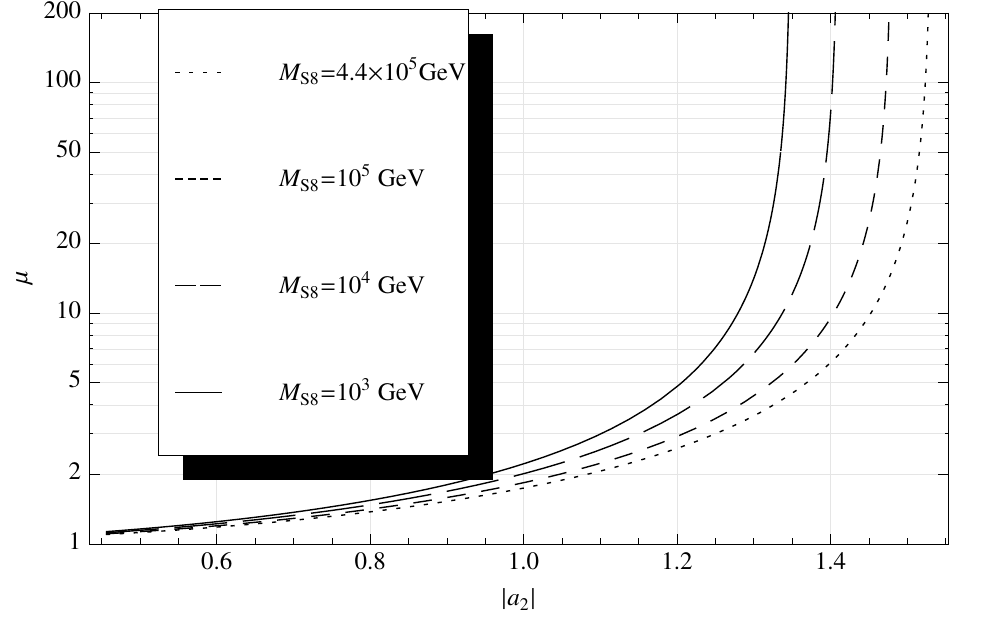}
  \includegraphics[width=.45\textwidth]{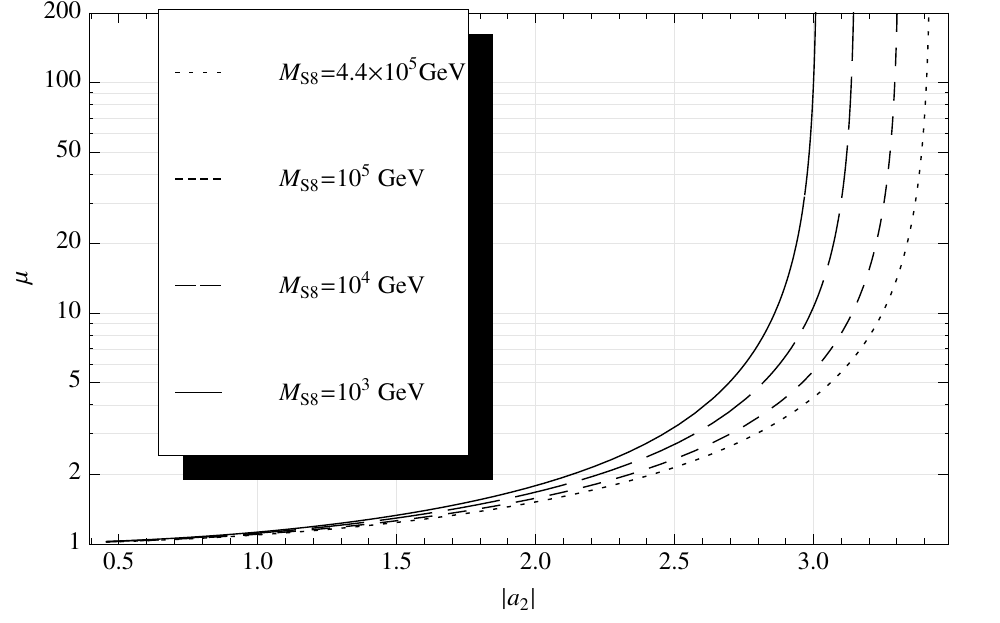}
  \caption{Left: $\mu$ versus $|a_2|$ in Adjoint $SU(5)$ model for $M_{\rho_3}=5\times10^{11}$~GeV, $M_{\rho_8}=10^{15}$~GeV and $\lambda_5=-0.5$;
  Right: same but for  $\lambda_5=-0.1$.}
\end{figure}


\section{Summary}\label{section:summary}

We considered the extensions of the SM by the heavy Majorana fermions, which govern the generation of both LG and neutrino masses, and divided them 
into 4 classes according to the different expressions for the $CP$ asymmetry in LG. Among them the 3 classes were discussed in the literature. However LG 
in the model with the lightest singlet and heavier triplets in the spectrum of the new fermions was not studied before. 
The bound on the $CP$ asymmetry in the hierarchical limit, which is known as Davidson--Ibarra bound, is discussed and explained how 
it can be relaxed in recently proposed freed LG. A particular realization of freed LG in Adjoint $SU(5)$ model is demonstrated. In addition, we pointed out the difference of the Davidson--Ibarra bound in the case 
with one massless neutrino from its standard form, which is valid for all massive neutrinos.
The possibility of non-hierarchical low energy LG, which is known as resonant LG, is also briefly discussed. 


\appendix

\section{$CP$ asymmetry in freed LG}\label{app:CPasymmetry}

The sum $\Sigma_\nu^\text{tree}$ in Eq.~\eqref{eq:Sigma} can be rewritten as
\begin{eqnarray}
 \Sigma_\nu^\text{tree} = \sum_j \frac{X_j}{M_j},
\end{eqnarray}
where 
\begin{eqnarray}
 X_j = \text{Im}[((YY^\dag)_{1j})^2].
\end{eqnarray}
By replacing in Eq.~\eqref{eq:Sigma} $M_\nu^\text{tree}$ from Eq.~\eqref{eq:nu_masses:1} by $M_\nu$ from Eq.~\eqref{eq:nu_masses:2} we get
\begin{eqnarray}
 \Sigma_\nu = \sum_j \frac{X_j}{M_j^\prime},
\end{eqnarray}
where 
\begin{eqnarray}
 M_j^\prime \equiv \left( \frac{1}{M_j}-\frac{a_j^2}{\Lambda} \right)^{-1},
\end{eqnarray}
which can be negative as well as positive.

For the case of only two new fermion mass states we have
\begin{eqnarray}
 \Sigma_\nu^\text{tree}=\mu\Sigma_\nu
\end{eqnarray}
with
\begin{eqnarray}
 \mu = \frac{M_2^\prime}{M_2} = \frac{\Lambda}{\Lambda-a_2^2M_2},
\end{eqnarray}
so that the $CP$ asymmetry in the hierarchical limit ${M_2/M_1\to\infty}$ takes the form~\cite{freedLG} 
\begin{eqnarray}\label{eq:epsilon_freed}
 \epsilon_1^{\text{freed}} = -\frac{3F}{16\pi} \mu \frac{M_1\Sigma_\nu}{(YY^\dag)_{11}}
\end{eqnarray}
with
\begin{eqnarray}
   \Sigma_\nu &=& \frac{1}{v_0^4} \sum_k m_k^2 \, {\rm Im} \left[ \left( Q^TD_\mathcal{M}^{1/2}R^\prime \right)_{1k}^2 \right], \\
    \left(YY^\dag\right)_{11} &=& \frac{1}{v_0^2} \sum_k m_k \left| (Q^T D_\mathcal{M}^{1/2}R^\prime)_{1k} \right|^2,
\end{eqnarray}
where the partly orthogonal matrix $R^\prime$ comes from the parametrization of $Y^\prime$ as
\begin{eqnarray}\label{eq:Casas-Ibarra^prime}
 Y^\prime = v_0^{-1}D_\mathcal{M}^{1/2}R^\prime D_\nu^{1/2}U^\dag,
\end{eqnarray}
similarly to Eq.~\eqref{eq:Casas-Ibarra}.

In the limit $M_1\ll \text{min}(\Lambda, M_2)$ we have~\cite{freedLG}
\begin{eqnarray}
 Q \approx \begin{pmatrix}
           1 & q \\
	   -q & 1
          \end{pmatrix}
\end{eqnarray}
with $q\approx \sin q\approx -a_1a_2 M_1 / \Lambda$, 
and the eigenvalues of the matrix $D_\mathcal{M}$
\begin{eqnarray} 
 \tilde M_1 \approx M_1 \left(1+a_1^2\frac{M_1}{\Lambda} \right), \qquad
\tilde M_2 \approx   M_2^\prime. 
\end{eqnarray}
Hence Eq.~\eqref{eq:epsilon_freed} can be rewritten as
\begin{eqnarray}
 \epsilon_1^{\text{freed}} \approx -\frac{3F}{16\pi}\, \frac{\mu M_1}{v_0^2} \,
\frac{ \text{Im}[\hat m_1] + q^2 r \, \text{Im}[\hat m_2] - 2q \, \text{Im}[\sqrt{r}\,\hat m_{12}]}{\tilde m_1 + q^2|r| \tilde m_2  - 2q\, \text{Re}[\sqrt{r}\,\tilde m_{12}]},
\end{eqnarray}
where $r=M_2^\prime/M_1$, and the light neutrino parameters are defined as
\begin{eqnarray}
 && \hat m_i = \sum_k m_k^2 (R_{ik}^{\prime})^2, \quad  \hat m_{12} = \sum_k m_k^2 R_{1k}^\prime R_{2k}^\prime,  \\
 && \tilde m_i = \sum_k m_k |R_{ik}^\prime|^2, \quad  \tilde m_{12} = \sum_k m_k R_{1k}^\prime R_{2k}^{\prime*}.
\end{eqnarray}
By taking the limiting cases $q^2\, |r|\ll1$ and $q^2\, |r|\gg1$ we get Eqs.~\eqref{eq:epsilon_freed1} and \eqref{eq:epsilon_freed2}, respectively.

\section*{{Acknowledgments}}

The author thanks Anatoly Borisov (MSU) for many useful discussions and comments on the ideas and the manuscript, Alexey Petrov and Gil Paz (WSU) for useful discussions.  
This work was supported in part by the EU ITN ``Unification in the LHC Era'',
contract PITN-GA-2009-237920 (UNILHC), by MIUR under contract 2006022501, and by the US Department of Energy under contract DE-SC0007983.


\end{document}